\documentclass{PoS}

\usepackage{lipsum}
\usepackage{color}
\usepackage{multicol}

\let\OLDthebibliography\thebibliography
\renewcommand\thebibliography[1]{
  \OLDthebibliography{#1}
  \setlength{\parskip}{0pt}
  \setlength{\itemsep}{0pt plus 0.3ex}
}

\title{Localizing the $\gamma$ rays from blazar PKS 1502+106}

\ShortTitle{Localizing the $\gamma$ rays from blazar PKS 1502+106}

\author{\speaker{Vassilis Karamanavis}\thanks{Member of the International Max
Planck Research School (IMPRS) for Astronomy and Astrophysics at the
Universities of Bonn and Cologne.} ~, L. Fuhrmann, T. P. Krichbaum, E.
Angelakis, J. Hodgson, ~~~~I. Myserlis, I. Nestoras, J. A. Zensus,\\
Max-Planck-Institut f\"{u}r Radioastronomie, Auf dem H\"{u}gel 69, D-53121
Bonn, Germany\\
E-mail: \email{vkaraman@mpifr.de}}
     
\author{H. Ungerechts and A. Sievers\\
Instituto de Radio Astronom\'{i}a Milim\'{e}trica, Avenida Divina Pastora 7,
Local 20, E-18012, Granada, Spain\\
}

\abstract{Blazars are among the most variable objects in the universe.
They feature energetic jets of plasma that launch from the cores of these
active galactic nuclei (AGN), triggering activity from radio up to
$\gamma$-ray energies. Spatial localization of the region of their MeV/GeV
emission is a key question in understanding the blazar phenomenon.

The flat spectrum radio quasar (FSRQ) PKS 1502+106 has exhibited extreme and
correlated, radio and high-energy activity that triggered intense monitoring by
the \textit{Fermi}-GST AGN Multi-frequency Monitoring Alliance (F-GAMMA) program
and the Global Millimeter VLBI Array (GMVA) down to $\lambda$3 mm (or 86 GHz),
enabling the sharpest view to date towards this extreme object.

Here, we report on preliminary results of our study of the $\gamma$-ray loud
blazar PKS 1502+106, combining VLBI and single dish data. We deduce the
critical aspect angle towards the source to be $\theta_{\rm c} = 2.6^{\circ}$,
calculate the apparent and intrinsic opening angles and constrain the distance
of the 86 GHz core from the base of the conical jet, directly from mm-VLBI but
also through a single dish relative timing analysis.

Finally, we conclude that $\gamma$ rays from PKS 1502+106 originate from a
region between $\sim$1--16$\,$pc away from the base of the
hypothesized conical jet, well beyond the bulk of broad-line region (BLR)
material of the source.}

\FullConference{12th European VLBI Network Symposium and Users Meeting,\\
		7--10 October 2014\\
		Cagliari, Italy}

\begin{document}

\section{Introduction}

Ever since $\gamma$ rays from AGN were detected, the exact location of 
their production site remains still unclear. It has been proposed
that high-energy emission may originate close to the central
engine---believed to be a supermassive black hole---within
the broad-line region (BLR), or several parsecs (pc) downstream the jet (cf.
\cite{2014MNRAS.441.1899F} and references therein).

PKS 1502+106 is a powerful blazar classified as a flat spectrum radio quasar
(FSRQ), with its central engine
having a mass of ${\sim} 10^{9}\,{\rm M}_{\odot}$ at redshift
$z=1.8385$\footnote{With ${\rm H}_{0} = 71 \, {\rm Km\,s}^{-1}\,{\rm Mpc}^{-1}$,
$\Omega_{\rm m} = 0.27$ and $\Omega_{\Lambda} = 0.73$ at $z=1.8385$, 1$\,$mas $=
8.53\,$pc.}\cite{2010ApJ...710..810A}. It underwent a period of pronounced
flaring activity starting early in 2008, that triggered high-cadence mm-VLBI and
filled-aperture, multi-wavelength observations. The wealth of available
single-dish data in addition to high-frequency Global mm-VLBI Array
(GMVA) observations at 86 (and 43) GHz evince that PKS 1502+106 is a unique case
towards a detailed understanding of flux density outbursts  and subsequent
structural variability. The powerful combination of single-dish monitoring and
mm-VLBI allows us to better constrain the $\gamma$-ray production site in PKS
1502+106.

\section{Observations and data reduction}

Our mm-VLBI observations with the
GMVA\footnote{\texttt{http://www3.mpifr-bonn.mpg.de/div/vlbi/globalmm}} comprise
6 epochs obtained between 2009 and 2012 with observations of PKS 1502+106 twice
per year. The following stations were included. Europe: Effelsberg (100-m),
Onsala (20-m), Pico Veleta (30-m), Plateau de Bure (6$\times$15-m),
Mets\"{a}hovi (14-m), Yebes (40-m); USA: all VLBA stations with 86 GHz
capability (8$\times$25-m). The data obtained in dual polarization mode were
correlated at the MPIfR. Uncalibrated data were analyzed with NRAO's AIPS. We
performed an \textit{a priori} visibility amplitude and phase calibration within
AIPS, using standard procedures \cite{2005AJ....130.1418J, 2012A&A...542A.107M}.
For subsequent imaging and analysis we used DIFMAP, where CLEANing and
visibility phase and amplitude self-calibration cycles were performed.

Single-dish data employed here were obtained within the framework of the F-GAMMA
program\footnote{
\texttt{http://www3.mpifr-bonn.mpg.de/div/vlbi/fgamma/fgamma.html}},
a coordinated effort for simultaneous observations at 11 bands in the
range between 2.64 to 345 GHz \cite{2007AIPC..921..249F,
2010arXiv1006.5610A, inest}. After initial flagging, pointing offset, opacity,
gain-curve, and sensitivity corrections were applied to the data.
From the broad F-GAMMA frequency coverage, here we employ the 86.24 GHz
band, matching our mm-VLBI monitoring. We also use data from the SMA
submillimeter calibrator list\footnote{
\texttt{http://sma1.sma.hawaii.edu/callist/callist.html}}
at 226.5 GHz \cite{2007ASPC..375..234G}.

\section{Results}

\begin{figure}
 \centering
  \includegraphics[width=\textwidth]{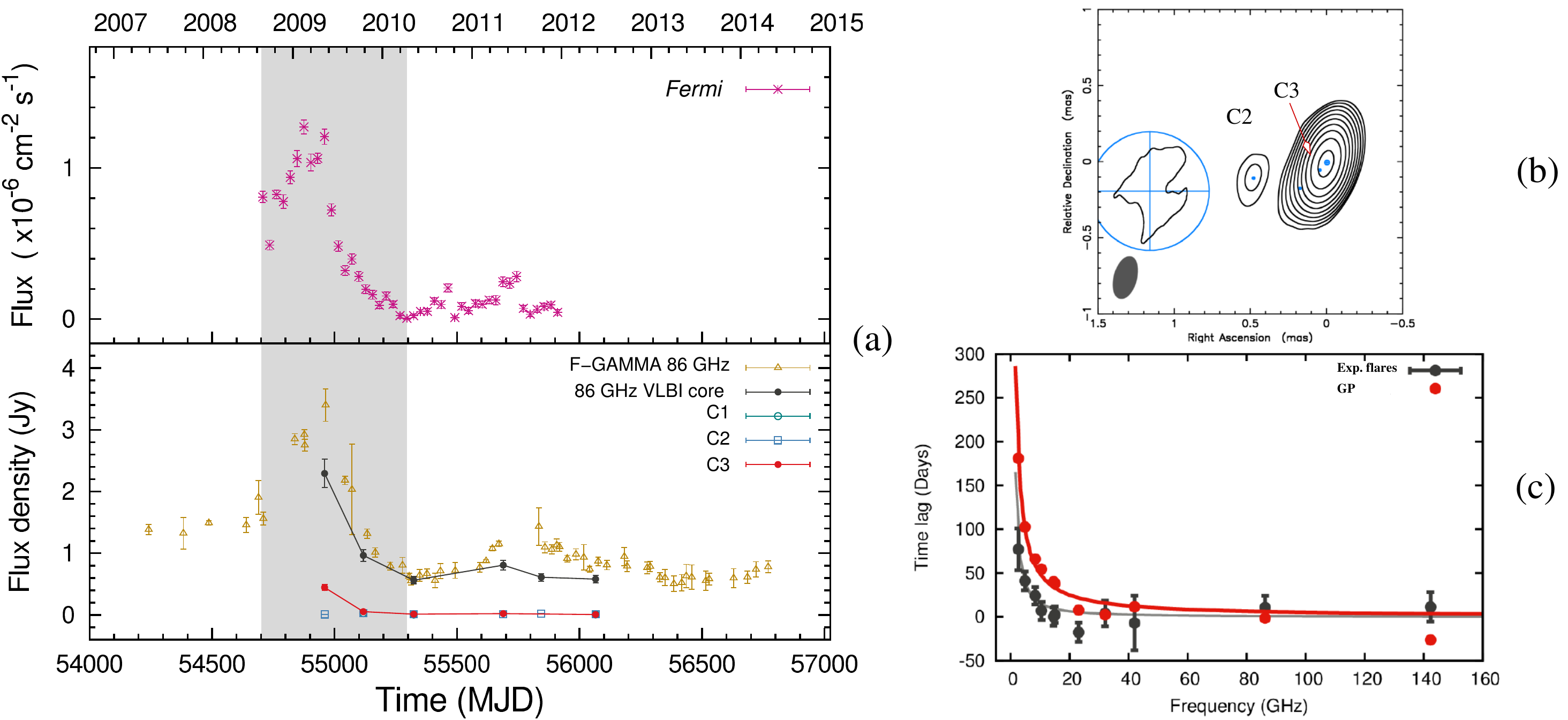}
  \caption{(a) Single dish total intensity and individual VLBI component
light curves at 86 GHz. \textit{Fermi} $\gamma$-ray data is shown additionally
(top), highlighting the correlated activity at different bands; (b) 86 GHz
modelfit map of PKS 1502+106 at epoch 2009.5. Contours are at 0.1\% to 68\% of
the peak flux density of 2.5 Jy/beam; (c) Frequency-dependent time delay for all
bands with respect to SMA 226.5 GHz data. Power law indices of $-1.3 \pm 0.3$
and $-1.0 \pm 0.1$ are obtained from exponential flare decomposition (black
curve) and Gaussian process regression (red curve), respectively.}
 \label{fig:1}
\end{figure}

\paragraph{Jet kinematics:}
As shown in Fig.$\,$\ref{fig:1}b, PKS 1502+106 exhibits a compact,
core-dominated morphology at 86 GHz with a one-sided parsec-scale jet, owing to
relativistic beaming. The jet can be decomposed into 3 modelfit components (C1
to C3). A linear fit to their radial separation from the core, yields their
apparent speeds within the jet flow. The most extreme superluminal motion of
about $22\,c$ at $15$ GHz\footnote{In addition to 43/86 GHz GMVA data, we have
reanalyzed VLBI data at 15 GHz obtained within the MOJAVE program
\cite{2009AJ....138.1874L}. These are not shown here; we only report the highest
speed of 22$\,c$ we obtain. The full VLBI analysis for PKS 1502+106 will appear
in a forthcoming article.} constrains the critical
viewing angle towards the source to be $\theta_{\rm c} = 2.6^{\circ}$.

\paragraph{Apparent and intrinsic jet opening angles:}
In estimating the apparent and intrinsic jet opening angles of PKS 1502+106, we
employ two methods. First, we use all modelfit components at all 3 available
frequencies (15, 43, and 86 GHz) and perform a single linear fit to their
deconvolved effective size with respect to radial separation from the core. In
the second method the opening angle is calculated for each resolved feature
separately, at each component's separation and according to the critical angle
constrained by its superluminal speed. Finally, we take the average as the
nominal opening angle for the jet. Resulting de-projected opening angles are
$\phi_{{\rm int}} = 1.5^{\circ}$ and $\phi_{{\rm int}} = 2.2^{\circ}$,
respectively. For a discussion on de-projecting blazar apparent opening angles
we refer to \cite{2013arXiv1301.5751C}.

\paragraph{Distance between the 3-mm VLBI core and the cone vertex:}
Under the assumption that the core takes up the entire jet cross section, the
de-projected distance of the 3-mm core from the vertex of a hypothesized
conical jet can be estimated by 
$d_{{\rm core}} = \left( 1.8~ \left< \mbox{FWHM} \right> _{{\rm core}} \right) /
\left[ 2~ \tan (\phi_{{\rm int}} / 2) \right]$.

With an average size for the core $\left< \mbox{FWHM}_{\rm core} \right>  =
0.03$ mas and a nominal opening angle between $\phi_{\rm int} =
2.2^{\circ}$--$1.5^{\circ}$, the 86 GHz core is placed ${\leq}$12--18$\,$pc
away from the vertex.

\paragraph{Single dish findings:}
Fig.$\,$\ref{fig:1}a shows the monthly-binned \textit{Fermi} $\gamma$-ray light
curve, the 3-mm F-GAMMA light curve along with those of VLBI knots C1, C2, C3
and the core. For the extraction of relevant
flare parameters we perform a fit with two methods. We employ a
decomposition into exponential flares and a non-parametric Gaussian
process (GP) regression. We fit each light curve separately, extracting the
flare amplitude $S_{\rm m}$, time scale and time of maximum flux density,
$t_{\rm m}$. Based on the different $t_{\rm m}$ per frequency we
gain insights into the opacity structure of the jet. The method
provides an alternative to VLBI core-shift measurements (cf.
\cite{2011MNRAS.415.1631K, 2014MNRAS.437.3396K}). Our relative timing analysis
for $S_{\rm m, 3\,mm}$ with respect to $S_{\rm m, 1\,mm}$ yields a time lag of
$\sim$10 days, translating into a distance between the single dish, 3-mm
$\tau=1$ surface to the vertex of the conical jet of ${\sim}$3 pc. Results of
our preliminary relative timing are shown in Fig.$\,$\ref{fig:1}c. Additionally,
a discrete cross-correlation function (DCCF) analysis between the 3\,mm and the
$\gamma$-ray light curves reveals that the radio flare lags behind its
high-energy counterpart by $14 \pm 11$ days, at a significance level above
$99\%$, thus placing the $\gamma$-ray production site $2.1\,$pc upstream of the
3-mm $\tau = 1$ surface (see \cite{2014MNRAS.441.1899F} for details).

\section{Conclusions and discussion}

We explored PKS 1502+106's extreme and significantly correlated variability in
radio and $\gamma$ rays employing both, single dish and VLBI methods. Our study
leads to the following preliminary conclusions:

1. PKS 1502+106 is a fast superluminal source, whose fastest $\beta_{\rm app}
\sim 22\,c$ constrains the critical aspect angle towards it, to
$\theta_{\rm c} = 2.6^{\circ}$. It features an intrinsic jet opening angle
$\phi_{{\rm int}} = 1.5^{\circ}$--$2.2^{\circ}$.

2. Based on VLBI images at $\lambda$3 mm, the core is situated
${\leq}$12--18$\,$pc away from the jet base.

3. The same calculation from single dish, relative timing analysis
yields a smaller value of ${\sim} 3\,{\rm pc}$. This is to be further
investigated. Possible explanations may include the fact that the VLBI core at
86 GHz is an unresolved feature, thus the distance reported above, represents
only an upper limit. The discrepancy could also hint that the filling factor in
the core region could be smaller. In any case, inferring $d_{\rm core}$ from
VLBI when the core is unresolved must be done with caution.

4. A DCCF analysis between the 3-mm and the $\gamma$-ray light curves places the
$\gamma$-ray production site at 2.1$\,$pc upstream of the 3-mm, single dish
$\tau = 1$ surface \cite{2014MNRAS.441.1899F}.

5. Even taking the extreme values for $d_{\rm core}$, coming from
single dish and VLBI (3 as opposed to 18$\,$pc) and combining with the DCCF
findings, we localize the $\gamma$ rays at ${\sim}$1--16$\,$pc away from the
base of the jet and at least the same distance from the black hole itself. In
both cases they originate downstream of the bulk BLR material in the
source ($R_{\rm BLR} \sim 0.1\,$pc \cite{2010ApJ...710..810A}) with important
consequences for the reservoir of target photons, available for inverse Compton
up-scattering and subsequent production of $\gamma$ rays.

\end{document}